\documentstyle[11pt]{article}\textheight 230mm\textwidth 150mm
            \newcommand{\be}{\begin{eqnarray}}
            \newcommand{\ee}{\end{eqnarray}}
            \newcommand{\eel}[1]{\label{#1}\end{eqnarray}}
\newcommand{\e}[1]{\label{e:#1}\end{eqnarray}}
     
            \newcommand{\ie}{{\em i.e.\ }}
            \newcommand{\ga}{{\gamma}}
 
            \newcommand{\la}{{\lambda}}
            
            \newcommand{\del}{{\delta}}

           \newcommand{\ra}{{\rightarrow}}
 \newcommand{\lea}{{\leftarrow}}

            \newcommand{\pet}{{\cal P}}

\newcommand{\ca}{{\cal C}}

            \newcommand{\beq}{\begin{quote}}
            \newcommand{\eq}{\end{quote}}
            \newcommand{\Om}{\Omega}
   
            \newcommand{\al}{\alpha}
            \newcommand{\ben}{\begin{enumerate}}
            \newcommand{\een}{\end{enumerate}}
            \newcommand{\bit}{\begin{itemize}}
            \newcommand{\ei}{\end{itemize}}
    	\newcommand{\nn}{\nonumber}
            \newcommand{\r}[1]{(\ref{e:#1})}
            \newcommand{\edfl}[1]{\label{#1}\end{df}}

\newcommand{\ve}{{\varepsilon}}

\newcommand{\dagg}{^{\dag}}

\newcommand{\bett}{{\bf 1}}
	
\def\d{\partial}
\def\cC{{\cal C}}

  \def\half{{1 \over 2}}

\def\JMP{{\sl J.\ Math.\ Phys.}}
\begin{document}
\begin{titlepage}
\noindent
G\"{o}teborg ITP 99-14\\
September 30, 1999

\vspace*{5 mm}
\vspace*{35mm}
\begin{center}{\LARGE\bf Open group  transformations\\
 within the Sp(2)-formalism}
\end{center} \vspace*{3 mm} \begin{center} \vspace*{3 mm}

\begin{center}Igor Batalin\footnote{On leave of absence from
P.N.Lebedev Physical Institute, 117924  Moscow, Russia\\E-mail:
batalin@td.lpi.ac.ru, batalin@sci.lebedev.ru.} and Robert
Marnelius\footnote{E-mail: tferm@fy.chalmers.se.}\\
\vspace*{7 mm} {\sl Department
of Theoretical Physics\\ Chalmers University of
Technology\\ G\"{o}teborg
University\\ S-412 96  G\"{o}teborg, Sweden}\end{center}
\vspace*{25 mm}
\begin{abstract}
Previously we have shown that open groups whose generators are in arbitrary
involutions may  be quantized within a ghost extended framework in terms of
the nilpotent BFV-BRST charge operator.
Here we show that they may also be quantized
within an Sp(2)-frame in which there are two odd anticommuting operators called
Sp(2)-charges. Previous results for finite open group transformations are
generalized to the Sp(2)-formalism. We show that in order to define open group
transformations on the whole ghost extended space we need Sp(2)-charges in the
nonminimal sector which contains dynamical Lagrange multipliers. We give an
Sp(2)-version of the quantum master equation with extended Sp(2)-charges and a
master charge of a more involved form,  which is proposed to represent the
integrability conditions of defining operators of connection operators and which
therefore should encode the  generalized quantum Maurer-Cartan equations for
arbitrary open groups. General solutions of this master equation are given in
explicit form. A further extended Sp(2)-formalism is proposed in which the group
parameters are quadrupled to a supersymmetric
set and from which all results may be
derived.
\end{abstract}\end{center}\end{titlepage}

\setcounter{page}{1}
\section{Introduction}
Open groups in which the generators are in
arbitrary involutions constitute a very general class of
continuous groups. The only framework in which one may hope to treat them
systematically  in quantum theory is within a ghost
extended BRST frame. This is because the quantum generators and their
algebra may
always be represented in terms of a BRST charge constructed according to the
BFV-prescription
\cite{BFV}. The quantization is then consistent if the corresponding BRST charge
operator is nilpotent, which always is possible to achieve for finite number of
degrees of freedom
\cite{BF}. In two recent papers we have developed the technique to
integrate these
arbitrary involutions and to construct finite group transformations within the
operator BFV-BRST frame
\cite{OG,OGT}. The most basic new equation we  found was a quantum master
equation involving an odd extended nilpotent BFV-BRST charge and an even  master
charge
\cite{OG}. This master equation is naturally defined in terms of the quantum
antibrackets defined in \cite{Quanti, GenQuanti}.  It encodes all integrability
conditions for defining operators of  the quantum connections, which in
turn were
shown to encode generalized Maurer-Cartan equations.  In
\cite{OGT} the formalism was further developed.
In particular we found an extended framework with
more ghosts from which all properties could be
extracted. Furthermore, we gave an explicit form of
the master charge that satisfies the master equation.

Since it is possible to embed generators in arbitrary involutions also into  two
anticommuting nilpotent odd operators, the BRST and the antiBRST charges
\cite{BaB}, there should exist a generalization of the
above results to this case. In gauge theories the  BRST and the antiBRST
formulation
 has been very useful and powerful. (There is also a corresponding
Lagrangian formulation of gauge theories which have been treated in many papers
\cite{Sp2}.) Here we shall consider an Sp(2)-version in which the BRST and
antiBRST
charges are cast into a charge with an Sp(2)-index.     A first attempt
to generalize the above results for open group transformations to  the
Sp(2)-formalism  was given in
\cite{Sp2QA}. There we showed that there
is an Sp(2)-valued connection operator whose
integrability conditions are determined by
an Sp(2)-valued quantum master equation in direct
analogy with the one of \cite{OG}. This
master equation was also shown to encode generalized
Maurer-Cartan equations. However, the resulting
finite group transformations were not defined on
the whole ghost extended space in this construction.
In this paper we propose a new more genuine
Sp(2)-version of
\cite{OG} in which the open group transformations
are defined on the entire extended space. As we
shall see this construction requires the
Sp(2)-charges to be given in the nonminimal
sector which also contains dynamical Lagrange multipliers. The results of the
present paper  allow us then to generalize
all properties given in \cite{OGT} to the
Sp(2)-formalism. We give a general Sp(2)-invariant solution of the master
equation,
and we find an extended Sp(2)-frame from which all results may be derived.

\section{Brief review of previous results.}
Let $\theta_{\al}$ with  Grassmann parities $\ve_{\al}(=0,1)$ be quantum
generators
of open group transformations.
They satisfy the commutator algebra
\be
 &&[\theta_{\al}, \theta_{\beta}]=i\hbar U_{\al\beta}^{\;\;\;\ga}\theta_{\ga},
\e{1}
where $U_{\al\beta}^{\;\;\;\ga}$ are operators in general.
Here and in the following we will only use graded commutators defined by
\be
&&[f, g]\equiv fg-gf(-1)^{\ve_f\ve_g}.
\e{05}
By means of additional ghost operators the algebra \r{1} may be embedded
into one
hermitian odd operator $\Om$ satisfying
\be
&&\half[\Om, \Om]=\Om^2=0.
\e{02}
$\Om$ is the BFV-BRST charge \cite{BFV} and this embedding
 is always possible to achieve for finite number of degrees of freedom
\cite{BF}.
 Explicitly $\Om$ is given by
\be
&&\Om=\ca^{\al}\theta_{\al}+\half\ca^{\beta}\ca^{\al}
U_{\al\beta}^{\;\;\ga}\pet_{\ga}(-1)^{\ve_{\beta}+\ve_{\ga}}+\cdots,
\e{2}
where $\ca^{\al}$ are ghost operators and $\pet_{\al}$ their conjugate
momenta with
Grassmann parities  $\ve(\ca^{\al})=
\ve(\pet_{\al})=\ve_{\al}+1$, satisfying the properties
 \be
&&[\ca^{\al},
\pet_{\beta}]=i\hbar\del^{\al}_{\beta},\quad(\ca^{\al})\dagg=\ca^{\al},
\quad\pet_{\al}\dagg=-(-1)^{\varepsilon_a}\pet_{\al}.
\e{04}
$\ca^{\al}$ and $\Om$  have ghost number one and $\pet_{\al}$ ghost number minus
one.   The $\Om$ in \r{2} is given in a $\ca\pet$-ordered form and the dots
denote
terms of higher powers in
$\ca^{\al}$ and
$\pet_{\al}$ which are determined by \r{02}. Notice that the hermiticity of
$\Om$
implies peculiar hermiticity properties of $\theta_{\al}$ in general.

In \cite{OG,OGT} we presented some new techniques to analyze finite open group
transformations within the above BRST framework. The results of this
analysis showed
that open group transformations on
the ghost extended space may be expressed in terms
of unitary operators of the form
\be
&&U(\phi)=\exp{\{-(i\hbar)^{-2}[\Om, \rho(\phi)]\}},
\e{3}
where $\phi^{\al}$, $\ve(\phi^{\al})=\ve_{\al}$,
are group parameters. $\rho(\phi)$ is an
hermitian odd operator
 with ghost number minus one, which was
required to satisfy $[\Om, \rho(0)]=0$ which
in turn yields
$U(0)=\bett$. A natural choice is
$\rho(\phi)\propto\pet_{\al}\phi^{\al}$ since $[\Om,
\pet_{\al}]$ represent the group generators in the extended BRST frame. The
latter
satisfy a closed algebra if the rank of the theory is zero or one.  For
rank two and
higher they only satisfy a closed algebra together with
$\pet_{\al}$. (The rank is equal to the maximal power of $\pet_{\al}$ in
$\Om$.)

In
terms of $U(\phi)$ we may define open group transformed operators $A(\phi)$ by
$A(\phi)\equiv U(\phi)AU^{-1}(\phi)$. It follows that $A(\phi)$ satisfies
the Lie
equations ($\d_{\al}\equiv\d/\d\phi^{\al}$)
\be
&&{A}(\phi)\stackrel{\lea}{{\nabla}}_{\al}\equiv
{A}(\phi)\stackrel{\lea}{\d_{\al}}
-(i\hbar)^{-1}[{A}(\phi),
{Y}_{\al}(\phi)]=0,
\e{301}
where
the quantum connection operator $Y_{\al}(\phi)$ is given by
\be
&&Y_{\al}(\phi)\equiv i\hbar U(\phi)
\left(U^{-1}(\phi)\stackrel{\lea}{\d_{\al}}\right).
\e{4}
The existence of a unitary operator $U(\phi)$
requires $Y_{\al}$ to satisfy the integrability
conditions
\be
&&Y_{\al}\stackrel{\lea}{\d_{\beta}}-Y_{\beta}
\stackrel{\lea}{\d_{\al}}(-1)^{\ve_{\al}\ve_{\beta}}=
(i\hbar)^{-1}[Y_{\al},
Y_{\beta}],
\e{5}
which also follow from the Lie equations \r{301}.
In order to have a representation of the form \r{3} the quantum connection
$Y_{\al}$ should from \r{4}  be of
the form
\be
&&Y_{\al}(\phi)=(i\hbar)^{-1}[\Om, \Om_{\al}(\phi)],
\quad\ve(\Om_{\al})=\ve_{\al}+1.
\e{6}
This was the starting point of ref.\cite{OG}.
$\Om_{\al}(\phi)$ has obviously ghost
number minus one. From \r{5} we found then
that the operators $\Om_{\al}(\phi)$ must
satisfy the integrability conditions
\be
&&\Om_{\al}\stackrel{\lea}{\d_{\beta}}-\Om_{\beta}
\stackrel{\lea}{\d_{\al}}(-1)^{\ve_{\al}\ve_{\beta}}-
(i\hbar)^{-2}(\Om_{\al}, \Om_{\beta})_{\Om}+\half
(i\hbar)^{-1}[\Om_{{\al}{\beta}}, \Om]=0,
\e{7}
where $\Om_{{\al}{\beta}}$ in general is a $\phi^{\al}$-dependent
 operator with ghost number minus
two and Grassmann parity
$\ve(\Om_{{\al}{\beta}})=\ve_{\al}+\ve_{\beta}$. The quantum
antibracket in \r{7} is defined by \cite{Quanti} (for their properties see
\cite{Quanti,GenQuanti})
\be
&&(f, g)_{\Om}\equiv\half \left([f, [{\Om}, g]]-[g, [{\Om},
f]](-1)^{(\ve_f+1)(\ve_g+1)}\right).
\e{8}
 From \r{7} one may then derive
integrability conditions for $\Om_{{\al}{\beta}}$ which
in turn introduces an operator $\Om_{{\al}{\beta}{\ga}}$
with ghost number minus three together with higher quantum antibrackets when the
$\Om$-commutator is divided out. Thus,  $Y_{\al}$ is
replaced by a whole set of operators, and  the
integrability conditions \r{5} for
$Y_{\al}$ is replaced by a whole set of
integrability conditions for these operators.
 These integrability conditions may be
viewed as generalized Maurer-Cartan
equations. The explicit form of $Y_{\al}$ and $\Om_{\al}$ are
\be
&&Y_{\al}(\phi)=\la^{\beta}_{\al}(\phi)
\theta_{\beta}(-1)^{\ve_{\al}+\ve_{\beta}}+\{\mbox{\small
possible ghost dependent terms}\}\nn\\
&&\Om_{\al}(\phi)=\la^{\beta}_{\al}(\phi)
\pet_{\beta}+\{\mbox{\small possible ghost
dependent terms}\},\quad
\la^{\beta}_{\al}(0)=\del^{\beta}_{\al},
\e{9}
where $\la^{\beta}_{\al}(\phi)$ are operators in general.  For
Lie group theories
$\la^{\beta}_{\al}(\phi)$ are pure functions and we may choose
$\Om_{\al}(\phi)=\la^{\beta}_{\al}(\phi)\pet_{\beta}$ and
$\Om_{{\al}{\beta}}=0$
in which case \r{7}
reduces to the standard Maurer-Cartan equations
\be
&&\d_{\al}{\la}_{\beta}^{\ga}-
\d_{\beta}{\la}_{\al}^{\ga}(-1)^{\ve_{\al}\ve_{\beta}}={\la}^{\eta}_{\al}
{\la}^{\del}_{\beta}
{U}^{\ga}_{{\del}{\eta}}(-1)^{\ve_{\beta}\ve_{\eta}
+\ve_{\ga}+\ve_{\del}+\ve_{\eta}}.
\e{10}

The crucial new discovery in \cite{OG} was that
the set of integrability conditions for
$\Om_{{\al}}$, $\Om_{{\al}{\beta}}$,  $\ldots$,
are encoded in a simple quantum master
equation involving an extended BRST charge $\Delta$ and a master charge $S$.
(This was proved to third order in $\eta^{\al}$.) It is
\be
&&(S,S)_{\Delta}=i\hbar[S, \Delta],
\e{11}
where the left-hand side is the antibracket \r{8} with $\Om$ replaced by
$\Delta$.
The odd operator $\Delta$ is defined
by
\be
&&\Delta\equiv\Om+\eta^{\al}\pi_{\al}(-1)^{\ve_{\al}},\quad \Delta^2=0,
\e{12}
where $\pi_{\al}$ is the conjugate momentum operator to $\phi^{\al}$ now turned
into an operator ($[\phi^{\al}, \pi_{\beta}]=i\hbar\del^{\al}_{\beta}$),
and where
$\eta^{\al}$,
$\ve(\eta^{\al})=\ve_{\al}+1$,  are new parameters which may be viewed as
superpartners to
$\phi^{\al}$.
The even operator $S$ is given by the following power expansion in $\eta^{\al}$
\be
&&S(\phi,
\eta)\equiv G+\eta^{\al}\Om_{\al}(\phi)+
\half\eta^{\beta}\eta^{\al}\Om_{{\al}{\beta}}(\phi)(-1)^{\ve_{\beta}}+
\nn\\&&+{1\over6}\eta^{\ga}\eta^{\beta}\eta^{\al}\Om_{{\al}{\beta}{\ga}}
(\phi)(-1)^{\ve_{\beta}+\ve_{\al}\ve_{\ga}}+
\mbox{\small terms of higher powers in $\eta^{\al}$},
\e{13}
where $G$ is the ghost charge operator. For
further details of this construction see
\cite{OG,OGT}.

\section{Previous Sp(2)-version}
In the Sp(2)-version of the BRST-theory the
quantum generators $\theta_{\al}$ in \r{1}
are embedded into two odd, hermitian  charge
operators $\Om^a$ where $a(=1,2)$ is an
Sp(2)-index.
$\Om^a$ have ghost number one and satisfy
\be
&&[\Om^a, \Om^b]=\Om^{\{a}\Om^{b\}}=0.
\e{15}
They  have the following $\ca\pet$-ordered form
\be
&&\Om^a=\ca^{\al a}\theta_\al+\half\ca^{\beta b}\ca^{\al a} {U}_{\al
\beta}^{\;\;\;\ga}\pet_{\ga b}(-1)^{\ve_\beta+\ve_\ga}+\cdots ,
\e{14}
where $\ca^{\al a}$ ($\ve(\ca^{\al a})=\ve_{\al}+1$) are Sp(2)-valued ghost
operators and $\pet_{\al a}$ their conjugate momenta satisfying
 \be
&&[\ca^{\al a},
\pet_{\beta b}]=i\hbar\del^{\al}_{\beta}
\del^a_b,\quad(\ca^{\al a})\dagg=\ca^{\al a},
\quad\pet_{\al a}\dagg=-(-1)^{\varepsilon_a}\pet_{\al a}.
\e{141}
The dots in \r{14}
denote terms  of higher orders in the ghost variables which are
 determined by \r{15}.

In \cite{Sp2QA} we proposed in analogy with
\r{301} and \r{6} that group transformed
operators $A(\phi)$ satisfy the Sp(2)-valued Lie equations
\be
&&{A}(\phi)\stackrel{\lea}{{\nabla}}^b_{\al a}\equiv
{A}(\phi)\stackrel{\lea}{\d_{\al}}\del^b_a
-(i\hbar)^{-1}[{A}(\phi),
{Y}^b_{\al a}(\phi)]=0,
\e{151}
where the
Sp(2)-valued connection
operators are of the form
\be
&&Y^b_{\al a}(\phi)=(i\hbar)^{-1}[\Om^b, \Om_{\al
a}(\phi)],\quad\ve(\Om_{\al a})=\ve_\al+1.
\e{16}
The analysis of these connections led to the result that
$Y_{\al}\equiv\half Y^a_{\al
a}$ may be viewed as a conventional connection operator like \r{4} while
\be
&&T^{ab}_\al(\phi)\equiv\ve^{\{ac}Y^{b\}}_{\al c}(\phi)
\e{17}
are constraint operators. The unitary operator
corresponding to $Y_{\al}$ only acts
within a ghost restricted subset of operators $A$ satisfying $[A,
T^{ab}_\al(\phi)]=0$. In a way this is natural
since the Sp(2)-formulation contains
more ghost variables then the conventional BRST frame.
By restricting the ghost
sector we essentially reduce the Sp(2)-formalism
to the standard one. Anyway within
this formalism we found that the integrability
conditions of $Y^b_{\al a}$ led to a
whole set of integrability conditions for
$\Om_{\al a}$ and higher operators
$\Om_{\al\beta a b}$, $\ldots$,  which were
symmetric in lower Sp(2)-indices and
which  was shown to  constitute generalized
Maurer-Cartan equations. This set of
equations were then shown  to be encoded in
the Sp(2)-valued quantum master equation
(proved to third order in
$\eta^{\al}$)
\be
&&(S, S)^a_{\Delta}=i\hbar[\Delta^a, S],
\e{18}
where the Sp(2)-valued quantum antibracket is defined by \cite{Quanti}
(their properties are given in \cite{Sp2QA,GenQuanti})
\be
&&(f, g)^a_{\Delta}\equiv\half \left([f, [\Delta^a, g]]-[g, [\Delta^a,
f]](-1)^{(\ve_f+1)(\ve_g+1)}\right).
\e{181}
The right-hand side is equal to
\r{8} with $\Om$ replaced by
$\Delta^a$. The two odd operators $\Delta^a$ were defined by (cf \r{12})
\be
&&\Delta^a\equiv\Om^a+j^a\eta^{\al} \pi_\al(-1)^{\ve_\al},\quad [\Delta^a,
\Delta^b]=0,
\e{19}
where again $\pi_{\al}$ is the conjugate momentum
to $\phi^{\al}$. $j^a$ are  even
Sp(2)-valued parameters. The even master charge
$S$ was in turn given by the following
power expansion in $\eta^{\al a}\equiv j^a\eta^{\al}$
\be
&&S(\phi,
\eta, j)\equiv G+j^a\eta^{\al}\Om_{\al
a}(\phi)+{1\over4} j^bj^a\eta^{\beta}
\eta^{\al}(-1)^{\ve_\beta}\Om_{\al\beta a
b}(\phi)+
\nn\\&&+{1\over36}j^cj^bj^a\eta^{\ga}\eta^{\beta
}\eta^{\al}(-1)^{\ve_\beta+\ve_\al\ve_\ga}
\Om_{\al \beta  \ga a b c}(\phi)+
\ldots.
\e{20}
The parameters $j^a$ provide for the  necessary Sp(2)-symmetrization in the
generalized Maurer-Cartan equations. For further details see
\cite{Sp2QA}.

\section{The new Sp(2)-proposal}
In this paper our object is to construct open group transformations within the
Sp(2)-formalism which are defined on the complete extended space. We expect then
that the group transformed operators $A(\phi)$
satisfy the Lie equations \r{301} and
that the quantum connections
$Y_{\al}$ are given by
\r{4} in terms of a unitary group element
$U(\phi)$. A natural generalization of \r{3} is  the
following Sp(2)-invariant expression
\be
&&U(\phi)=\exp{\{-(i\hbar)^{-2}\ve_{ab}[\Om^b, [\Om^a, R(\phi)]]\}},
\e{21}
where $R(\phi)$ is an even operator with ghost number minus two and such that
$U(0)=\bett$.
$\ve_{ab}$ is the Sp(2)-metric ($\ve_{ab}=-\ve_{ba}$,
$\ve^{ab}\ve_{bc}=\del^a_c$, $\ve^{12}=1$). We notice that commutators of
operators of the form $\ve_{ab}[\Delta^b,
[\Delta^a, A]]$ yield operators of the same
form, which means that $\ve_{ab}[\Delta^b, [\Delta^a, A]]$ is a natural
expression
for a group generator within the Sp(2)-formalism. If
$U(\phi)$ is of the form \r{21} then the quantum connection \r{4} may always be
written in the following Sp(2)-invariant form
\be
&&Y_{\al}=(i\hbar)^{-2}\half\ve_{ab}[\Om^b, [\Om^a, X_{\al}(\phi)]],
\e{22}
where $X_{\al}$ has Grassmann parity $\ve_{\al}$, the same as $Y_{\al}$. In this
way we  avoid the Sp(2)-valued connections \r{16} considered in \cite{Sp2QA}.
However, it remains to establish the existence
of quantum connections of this form. We
know that $Y_{\al}$ should satisfy the boundary conditions \r{9}. Now
this expression cannot be obtained from \r{22}
for any choice of $X_{\al}$ if $\Om^a$
are given in the minimal sector as in \r{14}. However, there is a
nonminimal sector involving dynamical Lagrange multipliers which does allow
for a
solution of the form
\r{22}. In this sector the $\Om^a$-operators have the form
\be
&&\Om^a=\ca^{\al a}\theta_\al+\half\ca^{\beta b}\ca^{\al a} {U}_{\al
\beta}^{\;\;\;\ga}\pet_{\ga
b}(-1)^{\ve_\beta+\ve_\ga}+\nn\\&&+\ve^{ab}\pet_{\beta b}\la^\beta+\half
\la^\beta\ca^{\al a} U^{\;\;\;\ga}_{\al\beta}\zeta_\ga+\cdots ,
\e{23}
where $\la^{\al}$, $\ve(\la^{\al})=\ve_{\al}$, are the Lagrange multipliers and
$\zeta_{\al}$ their conjugate momenta ($[\la^{\al},
\zeta_{\beta}]=i\hbar\del^{\al}_{\beta}$).
$\la^{\al}$ has ghost number
 two and $\zeta_{\al}$  minus two.
The expressions \r{23} are both $\ca\pet$- and
$\la\zeta$-ordered, and the higher order terms are determined by \r{15}. In the
nonminimal sector it is easily seen that the boundary conditions
 \r{9} for $Y_{\al}$  is reproduced by \r{22} with $X_{\al}$ given by
\be
&&X_{\al}(\phi)=-\la^{\beta}_{\al}(\phi)
\zeta_{\beta}(-1)^{\ve_{\al}+\ve_{\beta}}+\{\mbox{\small
possible ghost dependent terms}\}.
\e{24}
Thus, the quantum connections $Y_{\al}$
may have the Sp(2)-invariant form \r{22}
provided $\Om^a$ are given in the nonminimal sector with the form \r{23}. The
integrability conditions
\r{5} for
$Y_{\al}$ lead  through \r{22} to the following integrability conditions for
$X_{\al}$
\be
&&X_{\al}\stackrel{\lea}{\d_{\beta}}-X_{\beta}
\stackrel{\lea}{\d_{\al}}(-1)^{\ve_{\al}\ve_{\beta}}+
(i\hbar)^{-3}\half\{X_{\al},
X_{\beta}\}_{\Om}=(i\hbar)^{-1}[X_{\al\beta a}, \Om^a],
\e{25}
where the  operator $X_{\al\beta a}$ has ghost
number minus three and Grassmann parity
$\ve(X_{\al\beta a})=\ve_{\al}+\ve_{\beta}+1$, and   is in general
$\phi^{\al}$-dependent. In \r{25} we have introduced a
new quantum bracket defined by
\be
&&\{f, g\}_{\Om}\equiv \left[[f, \Om^a], \ve_{ab}[\Om^b, g]\right].
\e{26}
It has similar properties to the graded commutator \r{05} (see appendix A).
For Lie
group theories we may choose
$X_{\al}(\phi)=-\la^{\beta}_{\al}(\phi)
\zeta_{\beta}(-1)^{\ve_{\al}+\ve_{\beta}}$
where
$\la^{\beta}_{\al}(\phi)$ are functions in
which case \r{25} with $X_{\al\beta a}=0$
leads to the Maurer-Cartan equations \r{10}.

The integrability conditions for
$X_{\al\beta a}$ in \r{25} lead to higher operators whose integrability
conditions
lead to still higher operators etc. It is obviously a nontrivial task to invent
a master equation which encodes all these operators and their integrability
conditions. However, in the next section we will propose such a master
equation. To
make it easier to understand this proposal it should be helpful to view the
above
construction from
 the Sp(2)-treatment in \cite{Sp2QA}.

In \cite{Sp2QA} the quantum connections $Y_{\al}$ were given in terms of the
Sp(2)-valued connections \r{16} by
\be
&&Y_{\al}(\phi)=\half Y_{\al a}^a(\phi)=
\half(i\hbar)^{-1}[\Om^a, \Om_{\al a}(\phi)].
\e{27}
Obviously we may reproduce the expression \r{22} if we choose
\be
&&\Om_{\al a}(\phi)=\ve_{ba}(i\hbar)^{-1}[\Om^b, X_{\al}(\phi)].
\e{28}
We notice then that such a relation
 only is possible in the nonminimal sector.
It is easily seen that \r{28} with \r{23}
and
\r{24} yields
\be
&&\Om_{\al a}(\phi)=\la_{\al}^{\beta}(\phi)
\pet_{\beta a}+\{\mbox{\small possible ghost
dependent terms}\},
\e{29}
which are the boundary conditions imposed
in \cite{Sp2QA}. If we insert the expression
\r{28} into the Sp(2)-valued connections \r{16} we find
\be
&&Y_{\al a}^b(\phi)=\ve_{ca}(i\hbar)^{-2}[\Om^b, [\Om^c, X_{\al}(\phi)]].
\e{30}
By means of the Jacobi identities and \r{15} we find then that
\be
&&T_{\al}^{ab}(\phi)\equiv \ve^{\{ac}Y_{\al c}^{b\}}(\phi)=
-(i\hbar)^{-2}[\Om^{\{b}, [\Om^{a\}},
X_{\al}]]=0.
\e{31}
This implies that the choice \r{28} makes the Sp(2)-treatment in \cite{Sp2QA}
completely unconstrained and equivalent to our new proposal. Notice that
\r{31} means that $Y_{\al a}^b$ have only one
independent nonzero element. In fact, it
implies
\be
&&Y^1_{\al 2}=Y_{\al 1}^2=Y_{\al 1}^1-Y_{\al 2}^2=0\quad\Rightarrow\quad
Y_{\al}=Y_{\al 1}^1=Y_{\al 2}^2.
\e{32}
However, it is a nontrivial task to
implement the constraints \r{28} into the master
equation.

\section{The master equation}
We propose that the master equation
\be
&&(S, S)^a_{\Delta}=i\hbar[\Delta^a, S],
\e{33}
encodes all integrability conditions
starting with \r{25} for the operator $X_{\al}$
in our new proposal
\r{22} of the quantum connection operator $Y_{\al}$.
Although \r{33} is of the same
form as
\r{18} used in \cite{Sp2QA} the operators $\Delta^a$
and $S$ are here different from
the ones used there, which were \r{19} and \r{20}.
The two odd operators $\Delta^a$
we propose to have the following new form
\be
&&\Delta^a\equiv\Om^a+\eta^{\al a}
\pi_\al(-1)^{\ve_\al}+\rho^{\al}\xi_{\al
b}\ve^{ab}(-1)^{\ve_{\al}},\nn\\
&&[\Delta^a, \Delta^b]=\Delta^{\{a}\Delta^{b\}}=0,
\e{34}
where $\Om^a$ are the hermitian Sp(2)-charges in the nonminimal
sector  given by \r{23}.  $\eta^{\al a}$,
($\ve(\eta^{\al a})=\ve_{\al}+1$), are new operators and
$\xi_{\al a}$ their conjugate momenta ($[\eta^{\al a}, \xi_{\beta
b}]=i\hbar\del^{\al}_{\beta}\del^a_b$). $\rho^{\al}$,
($\ve(\rho^{\al})=\ve_{\al}$), are new parameters.
$\phi^{\al}$, $\rho^{\al}$ and
$\eta^{\al a}$ may be viewed as a superset
of new coordinates. The even master charge
$S$ has also a new form. Here  we propose it
to be given by a general power
expansion in $\eta^{\al a}$ and $\rho^{\al}$ (cf.\r{20}),
\be
&&S(\phi,\rho,\eta)=G+\eta^{\al a}\Om_{\al
a}(\phi)+\rho^{\al}\Om_{\al}(\phi)+\half\eta^{\beta b}\eta^{\al
a}(-1)^{\ve_{\beta}}\Om_{\al\beta
ab}(\phi)+\nn\\&&+\half\rho^{\beta}\rho^{\al}
\Om_{\al\beta}(\phi)+\rho^{\beta}\eta^{\al
a}\Om_{\al\beta a}(\phi)+\mbox{\small
terms of higher powers in $\rho^{\al}$ and
$\eta^{\al a}$},
\e{35}
where $G$ now is the ghost charge operator including the dynamical Lagrange
multipliers, \ie
\be
&&G\equiv-\half\biggl(\pet_{\al a}\cC^{\al a}-\cC^{\al a}\pet_{\al
a}(-1)^{\ve_\al}\biggr)-\biggl(\zeta_\al\la^\al+
\la^\al\zeta_\al(-1)^{\ve_\al}\biggr).
\e{36}
 The
Grassmann parities of the coefficient operators in \r{35}   are determined by the
ones of $\rho^{\al}$,
$\eta^{\al a}$ and that $S$ is even. Their
symmetry properties are
\be
&&\Om_{\al\beta}=\Om_{\beta\al}(-1)^{\ve_{\al}\ve_{\beta}}, \quad \Om_{\al\beta
ab}=-\Om_{\beta\al ba}(-1)^{\ve_{\al}\ve_{\beta}}, \quad\ldots
\e{361}
all determined by the $\rho\eta$-monomials
the coefficients are multiplied by in $S$.
Their ghost numbers
are determined by the following conditions:
$S$ is required to have total ghost number
zero and
$\Delta^a$ total ghost number one. The
latter requires $\eta^{\al a}$ to have ghost
number one which implies that $\xi_{\al a}$
has ghost number minus one which in turn
requires
$\rho^{\al}$ to have ghost number two.  One may notice that
\r{34} and
\r{35}  essentially reduce  to
\r{19} and
\r{20} if we set
$\rho^{\al}=0$ and $\eta^{\al a}=j^a\eta^{\al}$. However, if we first calculate
\r{33} and then take this limit then we will not get the same  equations as in
\cite{Sp2QA}. The expressions \r{34} and \r{35} yield
\be
&&[S, \Delta^a]=i\hbar\Om^a+\eta^{\beta b}[\Om_{\beta b},
\Om^a]+\rho^{\al}\biggl(-i\hbar\Om_{\al b}\ve^{ab}+[\Om_{\al}, \Om^a]\biggr)
+\nn\\&&+\half\eta^{\ga c}\eta^{\beta
b}(-1)^{\ve_{\ga}}\biggl(i\hbar\Om_{\beta b}\stackrel{\lea}{\d_{\ga}}\del^a_c
-i\hbar\Om_{\ga
c}\stackrel{\lea}{\d_{\beta}}\del^a_b
(-1)^{\ve_{\beta}\ve_{\ga}}+[\Om_{\beta\ga bc},
\Om^a]\biggr)+\nn\\&&+
\half \rho^{\ga}\rho^{\beta}\biggl(-i\hbar\Om_{\beta\ga
b}\ve^{ab}(-1)^{\ve_{\ga}}-i\hbar\Om_{\ga\beta
b}\ve^{ab}(-1)^{\ve_{\beta}({1+\ve_\ga})}+[\Om_{\beta\ga},
\Om^a]\biggr)+\nn\\&&+\rho^\ga\eta^{\beta
b}\biggl(i\hbar\d_{\beta}\Om_{\ga}\del^a_b(-1)^{\ve_{\ga}}-i\hbar\Om_{\beta\ga b
c}\ve^{ac}(-1)^{\ve_{\ga}}+[\Om_{\beta\ga b},
\Om^a]\biggr)+\nn\\&&+\mbox{\small terms
of higher powers in $\rho^{\al}$ and
$\eta^{\al a}$}.
\e{37}
Since the master equation \r{33} may equivalently be written as
\be
&&[S, [S, \Delta^a]]=i\hbar[S, \Delta^a]
\e{38}
it follows from \r{34} that
\be
&&[[S, \Delta^a], [S, \Delta^b]]=0.
\e{381}
From \r{38}
it is also straight-forward to calculate
\r{33} order by order in $\eta^{\al a}$ and
$\rho^{\al}$.  To the zeroth order and to
order $\eta^{\al a}$ the master equation
\r{33} is identically satisfied. Up to
second order we find the following nontrivial
equations:\\
({\small To order $\rho^{\al}$:})
\be
&&\Om_{\al
a}=\half(i\hbar)^{-1}\ve_{ab}[\Om_{\al}, \Om^b].
\e{39}
({\small To order $\eta^{\al a}\eta^{\beta b}$:})
\be
&& \Om_{\beta
b}\stackrel{\lea}{\d_{\ga}}\del^a_c-\Om_{\ga
c}\stackrel{\lea}{\d_{\beta}}\del^a_b(-1)^{\ve_{\beta}\ve_{\ga}}=
(i\hbar)^{-2}(\Om_{\beta
b}, \Om_{\ga c})_{\Om}^a-\half(i\hbar)^{-1}[\Om_{\beta\ga b c}, \Om^a].
\e{40}
({\small To order $\rho^{\al}\rho^{\beta}$:})
\be
&&\Om_{\al\beta
a}(-1)^{\ve_{\beta}}+\Om_{\beta\al
a}(-1)^{\ve_{\al}(\ve_{\beta}+1)}-{1\over4}
(i\hbar)^{-1}\biggl([\Om_{\al}, \Om_{\beta
a}]+[\Om_{\beta}, \Om_{\al a}](-1)^{\ve_{\al}\ve_{\beta}}\biggr)-\nn\\&&-
\half(i\hbar)^{-2}(-1)^{\ve_{\beta}}\ve_{ab}
(\Om_{\al},
\Om_{\beta})^b_{\Om}-{3\over4}(i\hbar)^{-1}\ve_{ab}[\Om_{\al\beta}, \Om^b]=0.
\e{41}
({\small To order $\rho^{\al}\eta^{\beta b}$:})
\be
&&\Om_{\al\beta
ab}=-\ve_{ab}\d_{\al}\Om_{\beta}+Z_{\al\beta ab},
\e{42}
where
\be
&Z_{\al\beta ab}\equiv&{1\over 3}(i\hbar)^{-1}[\Om_{\al a}, \Om_{\beta
b}](-1)^{\ve_{\beta}}-{2\over3}(i\hbar)^{-2}(\Om_{\al a},
\Om_{\beta})_{\Om}^c\ve_{cb}-\nn\\&&- {2\over3}(i\hbar)^{-1}[\Om_{\al\beta a},
\Om^c]\ve_{cb}(-1)^{\ve_{\beta}},\nn\\
\e{43}
where in turn the quantum Sp(2)-antibrackets are given by \r{181} with
$\Delta^a$
replaced by the nonminimal $\Om^a$.

Let us now analyze these equations. Eq.\r{39} determines $\Om_{\al a}$ in
terms of
$\Om_{\al}$. When compared to \r{28} it suggests that
\be
&&\Om_{\al}=2 X_{\al}(-1)^{\al}
\e{44}
 provided $\Om_{\al a}$ in $S$ may be
identified with $\Om_{\al a}$ in \r{16}. This
identification is possible since the
$bc$-symmetric parts of \r{40} agree exactly
with the integrability conditions for
$\Om_{\al a}$ in \cite{Sp2QA}.  Now \r{40}
seems to imply stronger conditions on
$\Om_{\al a}$, since the $bc$-antisymmetric
part of \r{40} is the nontrivial equation
\be
&&\ve^{ab}\biggl(\Om_{\beta
b}\stackrel{\lea}{\d_{\ga}}+\Om_{\ga
b}\stackrel{\lea}{\d_{\beta}}(-1)^{\ve_{\beta}\ve_{\ga}}\biggr)=
(i\hbar)^{-2}\ve^{cb}(\Om_{\beta
b}, \Om_{\ga c})_{\Om}^a-\half(i\hbar)^{-1}
[\ve^{cb}\Om_{\beta\ga b c}, \Om^a].\nn\\
\e{45}
However, we notice then first that \r{42} together with \r{361} imply
\be
&&\Om_{\al\beta
ab}=-\half\ve_{ab}\biggl(\d_{\al}\Om_{\beta}+
\d_{\beta}\Om_{\al}(-1)^{\ve_{\al}\ve_{\beta}}\biggr)+\half\biggl(Z_{\al\beta
ab}-Z_{\beta\al ba}(-1)^{\ve_{\al}\ve_{\beta}}\biggr),
\e{46}
which means that $\Om_{\al\beta ab}$ are completely expressed in terms of
$\Om$-operators with less indices.
By means of \r{46}, \r{39} and \r{41} we find then that \r{45} is identically
satisfied. This is related to the identical vanishing of $T_{\al}^{ab}(\phi)$ in
\r{31}. Thus, only the
$bc$-symmetric parts of
\r{40} are nontrivial.

Eq.\r{42} together with \r{361} imply
\be
&&\d_{\al}\Om_{\beta}-
\d_{\beta}\Om_{\al}(-1)^{\ve_{\al}\ve_{\beta}}=\half\ve^{ba}\biggl(Z_{\al\beta
ab}+Z_{\beta\al ba}(-1)^{\ve_{\al}\ve_{\beta}}\biggr),
\e{47}
 where the right-hand side  is
straight-forward to calculate from \r{43}.
We find  by means of \r{39} that it reduces
to
\be
&&\d_{\al}\Om_{\beta}-
\d_{\beta}\Om_{\al}(-1)^{\ve_{\al}\ve_{\beta}}={1\over4}(i\hbar)^{-3}\{\Om_{
\al},
\Om_{\beta}\}_{\Om}-{1\over12}(i\hbar)^{-3}\ve_{ab}[\Om^b, [\Om^a, [\Om_{\al},
\Om_{\beta}]]]- \nn\\&&- {1\over3}(i\hbar)^{-1}\biggl[(\Om_{\al\beta a}
(-1)^{\ve_{\beta}}-\Om_{\beta\al a}
(-1)^{\ve_{\al}(\ve_{\beta}+1)}),\Om^a\biggr].
\e{48}
These equations are identical to the integrability conditions \r{25} after the
identifications \r{44} and
\be
&&X_{\al\beta a}\equiv{1\over6}
\biggl(\Om_{\al\beta a}(-1)^{\ve_{\al}}-\Om_{\beta\al
a}(-1)^{\ve_{\beta}(\ve_{\al}+1)}+(i\hbar)^{-2}\ve_{ab}[[X_{\al},
X_{\beta}], \Om^b]\biggr).\nn\\
\e{49}
Notice that the combination of the
$\Om_{\al\beta a}$-operators entering $X_{\al\beta
a}$ are not determined by \r{41}. Eq.\r{48}
yields through \r{39} the $bc$-symmetric
parts of \r{40}.

To conclude we have found that the master
equation \r{33} with the ansatz \r{34} and
\r{35} for $\Delta^a$ and $S$ yields up to
second order in the new variables
$\eta^{\al a}$ and $\rho^{\al}$ exactly
the integrability conditions \r{25} of
$X_{\al}$ in our basic ansatz \r{22} for
the quantum connection operator $Y_{\al}$.
This convinces us that we will find the
integrability conditions of $X_{\al\beta a}$
in \r{49} at the third order in $\eta^{\al a}$
and $\rho^{\al}$ which in turn will
involve new
$X$-operators whose integrability
conditions are obtained at the fourth order and so
on exactly as we had in \cite{OG,OGT,Sp2QA}.

\section{Formal properties of the quantum master equation}
Consider the operators
$S(\al)$ and $\Delta^a(\al)$ defined by
\be
&&S(\al)\equiv e^{{i\over\hbar}\al F}
Se^{-{i\over\hbar}\al F}, \quad
\Delta^a(\al)\equiv e^{{i\over\hbar}\al F}
{\Delta^a} e^{-{i\over\hbar}\al F},
\e{501}
where $\al$ is a real parameter and
$F$  an arbitrary even operator. Obviously
\be
&&[\Delta^a(\al), \Delta^b(\al)]=
\Delta^{\{a}(\al)\Delta^{b\}}(\al)=0
\e{502}
and
\be
&&(S(\al), S(\al))^a_{\Delta(\al)}=
i\hbar[\Delta^a(\al), S(\al)],
\e{503}
provided $S$ satisfies the original
master equation \r{33}, and provided $\Delta^a$
is given by \r{34}. For
$F=S$ we have in particular
\be
&&S(\al)=S,\quad \Delta^a(\al)=
\Delta^a+(i\hbar)^{-1}[\Delta^a, S](1-e^{-\al}).
\e{504}
Thus, the master equation is satisfied if we replace $\Delta^a$ by
$\Delta^a+\beta(i\hbar)^{-1}[\Delta^a, S]$ for any real parameter
$\beta$.

If $F$ in \r{501} satisfies
\be
&&[\Delta^a, F]=0,
\e{505}
then $\Delta^a(\al)=\Delta^a$. In this
case $S(\al)$ is another solution of the
master equation \r{33} if $S$ is a given
solution.  In order for $S(\al)$ to have
total ghost number zero like $S$, $F$ must
have total ghost number zero, and in order
for $S(\al)$ to have
the same form \r{35} as $S$,
$F$ should not depend on $\pi_{\al}$ and
$\xi_{\al a}$. If we assume that
$F(\phi, \eta, \rho)$ may be expanded in
powers  of $\phi^{\al}$,
$\eta^{\al a}$ and $\rho^{\al}$ then the
solutions of \r{505} may be written as (the
proof is given in appendix B)
\be
&&F(\phi, \eta, \rho)=F(0,0,0)+\half
\ve_{ab}(i\hbar)^{-2}[\Delta^b, [\Delta^a,
\Phi(\phi,
\eta, \rho)]],
\e{506}
where
$\Phi$ is an even operator with total
ghost number minus two which does not depend on
$\pi_{\al}$ and $\xi_{\al a}$ and which satisfies
\be
\left.\ve_{ab}[\Delta^b,
[\Delta^a,
\Phi(\phi, \eta, \rho)]]\right|_{\phi=\eta=\rho=0}=0.
\e{5061}
 ($\Phi$ has the form
\r{c10} in appendix B.) $F(0,0,0)$ satisfies
$[\Om^a, F(0,0,0)]=0$. It is both
natural and consistent to impose the
restriction $F(0,0,0)=0$ in which case we find
the following class of invariance
transformations of the master equation \r{33}
\be
&&S\;\ra\;S'\equiv \exp{\biggl\{
-(i\hbar)^{-3}\half\ve_{ab}[\Delta^b,[\Delta^a,
\Phi]]\biggr\}}S\exp{\biggl\{(i\hbar)^{-3}
\half\ve_{ab}[\Delta^b,[\Delta^a,\Phi]]\biggr\}},
\e{507}
which leave the $\phi^{\al}$=$\eta^{\al a}$=$\rho^{\al}$=0
 component of $S$ invariant.
 This  was proposed to be the natural
automorphisms of the master equation in
\cite{Sp2QA}.
For the corresponding infinitesimal transformations we have
\be
&&\del S=(i\hbar)^{-3}[S,\half\ve_{ab}
[\Delta^b,[\Delta^a,\Phi]]],\nn\\
&&\del_{21}
S\equiv(\del_2\del_1-\del_1\del_2)S=(i\hbar)^{-3}
[S,\half\ve_{ab}[\Delta^b,[\Delta^a,\Phi_{21}]]],\nn\\
&&\Phi_{21}=(i\hbar)^{-3}\half\{\Phi_1, \Phi_2\}_{\Delta}.
\e{508}

We conjecture that the general solution of
the master equation \r{33} satisfying the
boundary condition $\left.S\right|_{\phi^{\al}=\eta^{\al
a}=\rho^{\al}=0}=G$ is given by
\be
S= \exp{\biggl\{
-(i\hbar)^{-3}\half\ve_{ab}[\Delta^b,[\Delta^a,
\Phi]]\biggr\}}G\exp{\biggl\{(i\hbar)^{-3}
\half\ve_{ab}[\Delta^b,[\Delta^a,\Phi]]\biggr\}},
\e{509}
where $\Phi(\phi,
\eta, \rho)$ has the same form as in \r{506}.
Notice that $S=G$ is a trivial solution of \r{33}.
The transformations \r{507} act transitively on \r{509}.

\section{Open group transformations within an extended Sp(2)-formalism}
The invariance transformations of the quantum
master equation \r{33} which follow
from the transformation formulas \r{501}
together with \r{505} suggest that we could
define extended group elements by (cf.\cite{OG})
\be
&&U(\phi, \eta, \rho)\equiv \exp{\{{i\over\hbar}
F(\phi,\eta, \rho)\}},\quad [\Delta^a, F]=0,\quad
[\tilde{G}, F]=0, \quad F(0,0,0)=0,
\e{601}
where $\tilde{G}$ is the extended ghost charge
\be
&&\tilde{G}\equiv G-\half\biggl(\xi_{\al a}\eta^{\al a}-\eta^{\al a}\xi_{\al
a}(-1)^{\ve_\al}\biggr)-\biggl(\sigma_\al\rho^\al+
\rho^\al\sigma_\al(-1)^{\ve_\al}\biggr),
\e{602}
where $G$ is given by \r{36}. $\sigma_{\al}$ are the conjugate
momenta  to
$\rho^{\al}$ which now are turned into operators ($[\rho^{\al},
\sigma_{\beta}]=i\hbar\del^{\al}_{\beta}$). Obviously
\be
&&[\tilde{G}, S]=0, \quad [\tilde{G}, \Delta^a]=i\hbar\Delta^a.
\e{603}
$F$ in \r{601} is according to appendix B given by (cf \r{506})
\be
&&F(\phi, \eta, \rho)=\half\ve_{ab}(i\hbar)^{-2}[\Delta^b, [\Delta^a,
\Phi(\phi,
\eta, \rho)]].
\e{604}
We notice that the $\eta^{\al a}$=$\rho^{\al}$=0
 component of \r{601} agrees with
\r{21} if we make the identification $R(\phi)=\Phi(\phi,0,0)$.

By means of \r{601}
we may  define extended group transformed
 operators by
\be
&&
\tilde{A}(\phi,
\eta, \rho)=U(\phi,
\eta, \rho){A}U^{-1}(\phi,
\eta, \rho),
\e{605}
where $A$ does not depend on $\phi^{\al}$,
$\eta^{\al a}$, $\rho^{\al}$ and their
conjugate momenta. At $\eta^{\al a}$=$\rho^{\al}$=0
  $\tilde{A}$ is the group
transformed  operators $A(\phi)$ satisfying the Lie
equations \r{301} with
the operator connections
\r{22}.  The
 operators $\tilde{A}$ in \r{605} satisfy the extended Lie equation
\be
&&\tilde{A}(\phi,\eta, \rho)\stackrel{\lea}{\tilde{\nabla}}_{\al}\equiv
\tilde{A}(\phi,\eta, \rho)\stackrel{\lea}{\d_{\al}}
-(i\hbar)^{-1}[\tilde{A}(\phi,\eta, \rho),
\tilde{Y}_{\al}(\phi,\eta, \rho)]=0,
\e{607}
where
\be
&&\tilde{Y}_{\al}(\phi,\eta, \rho)=i\hbar U(\phi, \eta, \rho)\left(U^{-1}(\phi,
\eta, \rho)\stackrel{\lea}{\d_{\al}}\right).
\e{608}
The expression \r{601} for $U(\phi, \eta, \rho)$
 implies that $\tilde{Y}_{\al}$ has
the form
\be
&&\tilde{Y}_{\al}(\phi,\eta, \rho)=
(i\hbar)^{-2}\half\ve_{ab}[\Delta^b,
[\Delta^a, \tilde{X}_{\al}(\phi,\eta, \rho)].
\e{609}
From our conjecture that  \r{509} is the general solution of
the master charge $S$ we see
that the master charge $S$ itself is a
group transformed ghost charge under the
extended group element \r{601} which
means that it satisfies the Lie equation
\be
&&S(\phi,\eta, \rho)\stackrel{\lea}{\tilde{\nabla}}_{\al}\equiv
S(\phi,\eta, \rho)\stackrel{\lea}{\d_{\al}}-
(i\hbar)^{-1}[S(\phi,\eta, \rho),
\tilde{Y}_{\al}(\phi,\eta, \rho)]=0.
\e{611}
This equation together with
 \r{609}
may be used to resolve $\tilde{X}_{\al}(\phi,\eta, \rho)$
 in terms of $S$ (cf the corresponding
treatment in section 6 of \cite{OG}).
\\ \\ \\
{\bf Acknowledgments}

I.A.B. would like to thank Lars Brink for
his very warm hospitality at the
Department of Theoretical Physics, Chalmers
and G\"oteborg University.   The work of
I.A.B. is  supported by INTAS grant 96-0308
 and by RFBR grants 99-01-00980,
99-02-17916. R. M. is supported by Swedish
Natural Science Research Council. \\ \\ \\
\def\theequation{\thesection.\arabic{equation}}
\setcounter{section}{1}
\renewcommand{\thesection}{\Alph{section}}
\setcounter{equation}{0}
\noindent
{\Large{\bf{Appendix A}}}\\ \\
{\bf Properties of the new quantum bracket \r{26}.}\\ \\
In \r{26} we introduced a new
quantum bracket which in a general algebraic framework
is defined by
\be
&&\{f, g\}_{Q}\equiv \left[[f, Q^a],
\ve_{ab} [Q^b, g]\right], \quad [Q^a, Q^b]=0,
\e{a1}
where  $f$ and $g$ are any
operators with  Grassmann parities $\ve(f)\equiv\ve_f$ and
$\ve(g)\equiv\ve_g$ respectively. $Q^a$, $a=1,2$,
are two odd anticommuting
operators. The commutators on the right-hand side
are  graded commutators defined by
\r{05}.
 The new quantum bracket \r{a1} has similar
properties to the graded commutator
\r{05}. Its properties are\\
\\ 1) Grassmann parity
\be
&&\ve(\{f, g\}_Q)=\ve_f+\ve_g.
\e{a3}
2) Antisymmetry
\be
&&\{f, g\}_Q=-\{g, f\}_Q(-1)^{\ve_f\ve_g}.
\e{a4}
3) Linearity
\be
&&\{f+g, h\}_Q=\{f, h\}_Q+\{g, h\}_Q, \quad (\mbox{for}\; \ve_f=\ve_g).
\e{a5}
4) If one entry is an odd/even parameter $\la$ we have
\be
&&\{f, \la\}_Q=0\quad {\rm for\ any\ operator}\;f.
\e{a6}
5) The generalized Jacobi identities
\be
&&\{f,\{g, h\}_Q\}_Q(-1)^{\ve_f\ve_h}+
cycle(f,g,h)=\nn\\&&=
\biggl(2[\{f, g\}_Q, \tilde{h}]+\{[\tilde{f}, g]+[f, \tilde{g}],
h\}_Q\biggr)(-1)^{\ve_f\ve_h}+
cycle(f,g,h),
\e{a7}
where the tilde operators are defined by
\be
&&\tilde{f}\equiv-\half\ve_{ab}[Q^b. [Q^a, f]].
\e{a8}
6) The generalized Leibniz rule
\be
&&\{f, gh\}_Q-\{f, g\}_Qh-g\{f, h\}_Q(-1)^{\ve_f\ve_g}=\nn\\
&&=\ve_{ab}[[f, Q^a], g][Q^b,
h](-1)^{\ve_g}+\ve_{ab}
[Q^b, g][[f,Q^a],h](-1)^{(\ve_f+1)(\ve_g+1)}=\nn\\&&=-[f, [g,
Q^a]\ve_{ab}[Q^b, h]]+[[f,g], Q^a]\ve_{ab}[Q^b, h]+[g, Q^a]\ve_{ab}[Q^b, [f,
h]](-1)^{\ve_f\ve_g}.
\e{a9}
The properties 1)-4) agree exactly with the
corresponding properties of the graded commutator
$[f,g]$ for arbitrary operators $f$ and $g$.
However, the graded commutator
satisfies  5) and 6) with zero on the
right-hand sides.

The new quantum bracket \r{a1}
 may also be expressed in terms of the quantum
Sp(2)-antibrackets. We find
\be
&&\{f, g\}_Q=\half\ve_{ab}
\biggl(([f, Q^a],g)_Q^b-(f, [Q^a, g])_Q^b\biggl)+{1\over
4}\ve_{ab}[Q^b, [Q^a, [f, g]]],
\e{a10}
where the antibrackets on the right-hand
side are defined by \r{181} with $\Delta^b$
replaced by $Q^b$.\\ \\ \\
\setcounter{section}{2}
\setcounter{equation}{0}
\noindent
{\Large{\bf{Appendix B}}}\\ \\
{\bf Proof of \r{506}.}\\ \\
Let  $F(\phi,\eta,\rho)$ be an operator
which does not depend on the conjugate momenta
to $\phi^{\al}$, $\eta^{\al a}$ and $\rho^{\al}$, \ie
$\pi_{\al}$,
$\xi_{\al a}$ and $\sigma_{\al}$. Let it
furthermore be expandable in powers of
$\phi^{\al}$, $\eta^{\al a}$ and $\rho^{\al}$.
 We want to solve the condition \r{505}, \ie
\be
&&[\Delta^a, F(\phi,\eta, \rho)]=0.
\e{c1}
To this purpose we introduce the operators
\be
&&\Lambda^a\equiv \eta^{\al a}\pi_{\al}
(-1)^{\ve_{\al}}+\rho^{\al}\xi_{\al
b}\ve^{ab}(-1)^{\ve_{\al}},\nn\\&&
\tilde{\Lambda}_a\equiv
\xi_{\al a}\phi^{\al}(-1)^{\ve_{\al}}+\sigma_{\al}\eta^{\al
b}\ve_{ba}(-1)^{\ve_{\al}},
\e{c2}
with the properties
\be
&&\ve(\Lambda^a)=\ve(\tilde{\Lambda}^a)=1,\quad
[\Lambda^a, \Lambda^b]=[\tilde{\Lambda}_a, \tilde{\Lambda}_b]=0.
\e{c3}
$\Lambda^a$ has total ghost charge plus one and
$\tilde{\Lambda}_a$ minus one.
We have then
\be
&&\Delta^a=\Om^a+\Lambda^a, \quad [\Delta^a,
\tilde{\Lambda}_b]=i\hbar\del^a_bN,
\nn\\&& N=\pi_{\al}\phi^{\al}+\xi_{\al a}\eta^{\al
a}+\sigma_{\al}\rho^{\al},\quad  [N, {\Lambda}^a]=
[N,\tilde{\Lambda}_a]=0.
\e{c4}
By commuting \r{c1} with $\tilde{\Lambda}_a$
and taking \r{c4} into account we
get
\be
&&i\hbar[N, F]=-\half[\Delta^a, [F, \tilde{\Lambda}_a]].
\e{c5}
This in turn implies
\be
&&(i\hbar)^2[N, [N, F]]=\half[\Delta^a,
[\Delta^b, [[F, \tilde{\Lambda}_b],
\tilde{\Lambda}_a]]={1\over 4}\ve_{ba}[\Delta^a,
[\Delta^b, [[F, \tilde{\Lambda}_d],
\tilde{\Lambda}_c]]]\ve^{cd},
\e{c6}
where the last equality follows from
$[\Delta^a, \Delta^b]=0$. Now
\be
&&(i\hbar)^{-1}[N, F(\phi,\eta, \rho)]=
-\biggl(\phi^a{\d\over\d\phi^a} +
\eta^a{\d\over\d\eta^a}+\rho^{\al}
{\d\over\d\rho^a}\biggr) F(\phi,\eta, \rho)
\e{c7}
which equivalently may be written as
\be
&&(i\hbar)^{-1}[N, F(\al\phi, \al\eta, \al\rho)]
=-\al{d\over d\al}F(\al\phi,\al\eta,
\al\rho)
\e{c8}
where $\al$ is a real parameter. From this
expression it follows that the solution of
\r{c6} may be written as
\be
&&F(\phi, \eta, \rho)=F(0, 0, 0)+\nn\\&&+{1\over
4}(i\hbar)^{-2}\int_0^1{d\al\over\al}
\int_0^{\al}{d\beta\over\beta}\ve_{ba}
[\Delta^a, [\Delta^b, [[F(\beta\phi,
\beta\eta, \beta\rho), \tilde{\Lambda}_d],
\tilde{\Lambda}_c]]]\ve^{cd}\equiv \nn\\&&\equiv F(0, 0,
0)+(i\hbar)^{-2}\half\ve_{ab}[\Delta^b, [\Delta^a,
\Phi]] ,
\e{c9}
where $F(0,0,0)$ satisfies
\be
&&[\Delta^a, F(0,0,0)]=[\Om^a, F(0,0,0)]=0.
\e{c91}
Eq.\r{c9} is the assertion in \r{506}. Notice that
 $\Phi$ has the explicit form
\be
&&\Phi(\phi,\eta,
\rho)=\half(i\hbar)^{-2}\int_0^1{d\al\over\al}
\int_0^{\al}{d\beta\over\beta}[[F(\beta\phi,
\beta\eta, \beta\rho),
\tilde{\Lambda}_d],
\tilde{\Lambda}_c]\ve^{cd} +\nn\\&&+(i\hbar)^{-1}
[\Delta^a,  K_a(\phi,\eta, \rho)],
\e{c10}
where $K_a(\phi,\eta, \rho)$ are arbitrary
odd operators with total ghost number minus three
 which do not
depend on
$\pi_{\al}$,
$\xi_{\al a}$ and $\sigma_{\al}$.

A similar analysis was performed within
the Sp(2)-version of the BV-quantization in
\cite{BT}.


\begin{thebibliography}{Simple}



\bibitem{BFV}I. A. Batalin and G. A Vilkovisky,
\ {\sl  Phys. Lett.}
\ {\bf B69},\ 309\ (1977),\\
 E. S. Fradkin T. E. Fradkina, \ {\sl  Phys. Lett.}
\ {\bf B72},\ 343\ (1978),\\
I. A. Batalin and E. S. Fradkin, \ {\sl  Phys. Lett.}
\ {\bf B122},\ 157\ (1983).

\bibitem{BF}I. A. Batalin and E. S. Fradkin,
\ {\sl Phys. Lett.}\ {\bf B128},\ 303\
(1983);\\ {\sl Riv. Nuovo Cim.}\ {\bf 9},\ 1\
(1986);\ {\sl Ann. Inst. Henri Poincar\'{e}}\
{\bf 49},\ 145\ (1988).

\bibitem{OG}I. A. Batalin and
R. Marnelius,
\ {\sl Phys.
Lett.}\ {\bf B441},\ 243 \ (1998).

\bibitem{OGT}I. A. Batalin and
R. Marnelius, \ {\sl Mod. Phys. Lett.}\ {\bf A14},\ 1643\ (1999).

\bibitem{Quanti}I. A. Batalin and
R. Marnelius,  \
{\sl Phys. Lett.}\ {\bf B434},\ 312\ (1998).

\bibitem{GenQuanti}I. A. Batalin and
R. Marnelius,  \
{\em General quantum antibrackets},\ {\tt hep-th/9905083}.\\
(To be published in {\sl Theor. Math. Phys.})

\bibitem{BaB}G. Curci and R. Ferrari, \
{\sl Nuovo Cimento}\ {\bf A32},\ 151\ (1976),\\
I. Ojima, \
{\sl Progr. Theor. Phys. Lett.}\ {\bf 64},\ 625\ (1980),\\
N. Nakanishi and I. Ojima, \
{\sl Z. Phys. Lett.}\ {\bf C6},\ 155\ (1980),\\
L. Bonora and M. Tonin, \ {\sl Phys. Lett.}\ {\bf B98},\ 48\ (1981),\\
R. Delbourgo and P. D. Jarvis, \ {\sl J. Phys.}, {\bf A15}, 611 (1982),\\
S. Hwang, \
{\sl Nucl. Phys.}\ {\bf B231},\ 386\ (1984); {\bf B322},\ 107\ (1989),\\
V. P. Spiridonov, \ {\sl Nucl. Phys.}\ {\bf B308},\ 527\ (1988),\\
I.A.~Batalin, P.M.~Lavrov, and I.V.~Tyutin,
\JMP\ {\bf 31}, \ 6, \ 2708 (1990).\\
P.~Gr\'{e}goire and M.~Henneaux, \ {\sl  Phys. Lett.}\
{\bf B277}, \   372\ (1993);\\
{\sl Commun. Math. Phys.}\
{\bf 157}, \   279\ (1993).

\bibitem{Sp2}
I.A.~Batalin, P.M.~Lavrov, and I.V.~Tyutin,
\JMP\ {\bf 31}, 1487 (1990); {\bf 32}, 532, 2513 (1990),\\
C.M.~Hull,  {\sl Int. J. Mod. Phys}, {\bf A5},  1871 (1990),\\
M.~Henneaux, \  {\sl Phys. Lett.} {\bf B282},  372 (1992),\\
G.~Barnich, R.~Constantinescu and P.~Gr\'{e}goire, \ {\sl
Phys. Lett.} {\bf B293},   353 (1992),\\
P.~Gr\'{e}goire and M.~Henneaux, \ {\sl J. Phys.}
{\bf A26},   6073 (1993),\\
P.H.~Damgaard and F.~De~Jonghe, \ {\sl Phys. Lett.}
{\bf B305},  59 (1993).\\
I. A. Batalin and
R. Marnelius,  \
{\sl Phys. Lett.}\ {\bf B350},\ 44\ (1995),\\
I. A. Batalin,
R. Marnelius and A.M.~Semikhatov, \
{\sl Nucl. Phys.}\ {\bf B446},\ 249\ (1995),\\
A.~Nersessian and P.H.~Damgaard,  \
{\sl Phys. Lett.}\ {\bf B355},\ 150\ (1995),\\
I. A. Batalin and
R. Marnelius,  \
{\sl Nucl. Phys.}\ {\bf B465},\ 521\ (1996)\\
M.A. Grigoriev and A.M.~Semikhatov,  \
{\sl Phys. Lett.}\ {\bf B417},\ 259\ (1998);\\
{\em A Kahler structure of
triplectic geometry}, \ {\tt hep-th/9807023}.




\bibitem{Sp2QA}I. A. Batalin and
R. Marnelius, \ {\sl Nucl. Phys.}\ {\bf B551},\ 450 \ (1999).



\bibitem{BT}I. A. Batalin and
I. V. Tyutin, \ {\sl Theor. Math.  Phys.}\ {\bf 114},\ 198 \ (1998).

\end{thebibliography}
\end{document}